\documentclass[12pt,a4paper]{article}
\usepackage{amsmath}

\def\,{\ifmmode\mskip\thinmuskip\else\leavevmode\thinspace\fi}

\title {General solution of asymptotic conditions for electromagnetic form factors of
hadrons represented by VMD model}

\author{Stanislav~Dubni\v{c}ka$^1${\footnote{e-mail: fyzidubn@savba.sk}}, Anna Zuzana~Dubni\v{c}kov\'a$^2${\footnote{e-mail: dubnickova@fmph.uniba.sk}},
 \\ and  Peter Weisenpacher$^3${\footnote{e-mail: upsyweis@savba.sk}}}
\date{\empty}

\begin{document}
\maketitle
\begin{center} {
$^{1}$ \it Inst. of Physics, Slovak Academy of Sciences,  Bratislava,
Slovak Republic \\
$^{2}$ \it Dept. of Theor. Physics, Comenius Univ., Bratislava,
Slovak Republic \\
$^{3}$ \it Inst. of Informatics, Slovak Academy of Sciences, Bratislava,
Slovak Republic }\\
\end{center}

\vspace*{0.5cm}

\begin{abstract}
General solution of asymptotic conditions, derived previously for
$n$ vector-meson parametrization of electromagnetic form factor of
any strongly interacting particle with the asymptotics
$\sim_{|t|\to\infty} t^{-m} (m<n)$ and combined with a form factor
normalization condition, is presented. The special case of $m=n$
and the solution of asymptotic conditions without any form factor
normalization are discussed too.
\end{abstract}

\newpage

\section{Introduction}

Recently, starting with different properties of the
electromagnetic (EM) form factor (FF) $F_h(t)$ of a strongly
interacting particle to be saturated by $n$ vector-mesons and
possessing the asymptotic behaviour $\sim_{|t|\to\infty} t^{-m} \;
(m\leq n)$, two dissimilar systems of $(m-1)$ linear homogenous
algebraic equations for coupling constant ratios of vector-mesons
to hadron under consideration were derived \cite{dub1}. Though
they  really look differently, in \cite{dub1} it has been
demonstrated explicitly that both systems are exactly equivalent.

In this paper we are concerned with a more simple one, derived by
means of the superconvergent sum rules for the imaginary part of the
EM FF, in which the coefficients are simply even powers of
the corresponding vector-meson masses. In more detail, we look for its
general solution, which leads to the VMD representation of $F_h(t)$ with
the required asymptotics.

There are interesting three cases appearing in  various physical
situations, and all of them are discussed in this paper.

The first one appears in the construction of the unitary and
analytic model of EM structure  \cite{dub2}  of any
strongly interacting particle with a number of building quarks $n_q
>2$, when at the first stage one has need for the VMD
parametrization of FF under consideration with
the required asymptotics and normalization. The latter is found by a
combination of  the  $(m-1)$ asymptotic conditions with the FF normalization
condition and by a general
solution of the obtained  $m$ linear algebraic equations for  $n$
coupling constant ratios. As a result, the FF depends then
on  the  $(n-m)$ coupling constant ratios as free parameters of the
model.

The second case is obtained from the previous one for $m\equiv n$ and it leads
to expressions of all coupling constant ratios through the vector-meson
masses. If the latter are known, numerical values of the coupling constant
ratios are found, like in \cite{dub3}, for tensor coupling constants of
vector-mesons to nucleons.

The third case appears naturally in the determination of strange
FF behaviours of strongly interacting particles with the spin
$s>0$ from the isoscalar parts of the corresponding  EM FF's.
For instance, the value of the strangeness nucleon
magnetic moment $\mu_s$ is unknown in advance and thus, the
corresponding strange magnetic FF (as a consequence also the
strange Pauli FF's) model is constructed without the
normalization \cite{dub4}. In order to keep some inner analytic
structure of the corresponding EM form factor model,
one has to construct it also without any normalization, though in
the electromagnetic case it is exactly known experimentally to be
equal to the magnetic moment of the nucleon. So  in such a
situation one has to solve the asymptotic conditions in the form
of $(m-1)$ linear homogeneous algebraic equations for $n$ coupling
constant ratios.  The resultant solutions express the  $(m-1)$ coupling
constant ratios through the rest $(n-m+1)$ ones  which are then
free parameters of the model.

More details about the general solutions of asymptotic conditions and their
consequences for all three specific cases can be found in the next section.
The last section is devoted to conclusions and discussion.

\section{General solution of asymptotic conditions}

First, we  look for a general solution of the asymptotic conditions to
be combined with the FF norm  when FF is saturated by
more vector-meson resonances than the  power determining the FF
asymptotics.

If we assume that EM FF of any strongly interacting particle is
well approximated by a finite number $n$ of vector-meson exchange
tree Feynman diagrams,  one finds the VMD pole
parametrization
\begin{equation}
F_h(t)=\sum_{i=1}^n\frac{m_i^2}{m_i^2-t}(f_{ihh}/f_i) \label{ro1}
\end{equation}
where $t=-Q^2$ is the momentum transfer squared of the virtual photon, $m_i$
are the masses of vector mesons, and $f_{ihh}$ and $f_i$ are the coupling
constants of the vector-meson to hadron and vector-meson-photon transition,
respectively. Furthermore, let us assume that EM FF in (\ref{ro1}) has the
asymptotic behaviour
\begin{equation}
F_h(t)_{|t|\to\infty} \sim t^{-m} \label{ro2}
\end{equation}
and it is normalized at $t=0$ as follows:
\begin{equation}
F_h(0)=F_0. \label{ro3}
\end{equation}

The requirement for the conditions (\ref{ro3}) and (\ref{ro2}) to be satisfied
by (\ref{ro1}) (including also the results of ref. \cite{dub1}) leads to the
following system of $m$ linear algebraic equations:
\begin{eqnarray}
\sum_{i=1}^n a_i&=&F_0  \label{ro4} \\
\sum_{i=1}^n m_i^{2r}a_i&=&0,  \;\;\; r=1,2,...,m-1 \nonumber
\end{eqnarray}
for $n$ coupling constant ratios $a_i=(f_{ihh}/f_i)$. Therefore,  a solution
of (\ref{ro4}) will be looked for  $m$ unknowns $a_1$,...,$a_m$ and
$a_{m+1}$,...,$a_n$ will be considered as free parameters of the model. Then, the
system (\ref{ro4}) can be rewritten in the matrix form
\begin{equation}
\mathbf{M}\mathbf{a}=\mathbf{b}, \label{ro5}
\end{equation}
with the $m\times m$ Vandermonde matrix $\mathbf{M}$
\begin{equation}
\mathbf{M}=\begin{pmatrix}
1 & 1 & \hdots & 1\\
m_1^2 & m_2^2 & \hdots & m_m^2 \\
m_1^4 & m_2^4 & \hdots & m_m^4 \\
\hdotsfor 4 \\
m_1^{2(m-1)} & m_2^{2(m-1)} & \hdots &  m_m^{2(m-1)} \\
\end{pmatrix}\qquad                                        \label{ro6}
\end{equation}
and the column vectors
\begin{eqnarray}
\mathbf{a}=\begin{pmatrix} a_1\\a_2 \\a_3 \\
\vdots\\a_m\\\end{pmatrix},\qquad \;\;\;
\mathbf{b}=\begin{pmatrix}
F_0-\sum_{k=m+1}^na_k\\-\sum_{k=m+1}^nm^2_k a_k \\
-\sum_{k=m+1}^nm^4_k a_k
\\ \hdotsfor 1\\ -\sum_{k=m+1}^nm^{2(m-1)}_k a_k\\
\end{pmatrix}. \qquad
\label{ro7}
\end{eqnarray}
The Vandermonde determinant of the matrix (\ref{ro6}) is different from zero
\begin{equation}
det\mathbf{M}=\prod^m_{\substack{j,l=1,\\ j<l }} (m_l^2-m_j^2).
\label{ro8}
\end{equation}
This has been proved explicitly by reducing the matrix (\ref{ro6})
to the triangular form and then taking into account the fact  that the
determinant of a triangular matrix is the product of its main
diagonal elements.

As a consequence of (\ref{ro8}) a nontrivial solution of
(\ref{ro5}) exists. To find  the latter we use Cramer's
Rule despite  the fact that computationally Cramer's Rule for
$m>3$ offers no advantages over the Gaussian elimination method.
However, in our case (as one can see further) all calculations are
for the most part reduced  to a calculation of the
Vandermonde type determinants,  and there is no problem to come to
the explicit solutions.

So, the corresponding solutions of (\ref{ro5}) for $i=1,...,m$ are
\begin{equation}
a_i=\frac{det\mathbf{M_i}}{det\mathbf{M}} \label{ro9}
\end{equation}
where the matrix $\mathbf{M_i}$ takes the following form
\begin{equation}
\mathbf{M_i}=\begin{pmatrix} 1   \hdots & 1 & F_0- \sum_{\substack
{k=m+1}}^na_k & 1 & \hdots & 1\\ m_1^2   \hdots & m_{i-1}^2 & 0-
\sum_{\substack {k=m+1}}^n m_k^2a_k & m_{i+1}^2 & \hdots &
m_m^2 \\ m_1^4  \hdots & m_{i-1}^4 & 0-
\sum_{\substack {k=m+1}}^n m_k^4a_k& m_{i+1}^4  & \hdots & m_m^4 \\ \hdotsfor 6\\
m_1^{2(m-2)}  \hdots  & m_{i-1}^{2(m-2)} &  0- \sum_{\substack {k=m+1}}^nm_k^{2(m-2)}a_k & m_{i+1}^{2(m-2)} & \hdots & m_m^{2(m-2)} \\
m_1^{2(m-1)}   \hdots & m_{i-1}^4  &  0-
\sum_{\substack {k=m+1}}^nm_k^{2(m-1)}a_k & m_{i+1}^{2(m-1)} & \hdots &  m_m^{2(m-1)} \\
\end{pmatrix}.\qquad           \label{ro10}
\end{equation}
Since any determinant is an additive function of each column,
 for each scalar $C$,
$det(A_1,...,CA_i,...A_n)$=$Cdet(A_1,...A_i,...A_n)$\\ and
$det(A_1,..,A_{i-1},\sum_k x_kA_k,A_{i+1},...,A_n)$=$\sum_kx_k
det(A_1,...,A_{i-1},A_k,A_{i+1},...,A_n)$. As a result, for a determinant of
the matrix $\mathbf{M_i}$ one can write the decomposition
\begin{eqnarray}
& &det\mathbf{M_i}= \begin{vmatrix}
1 & 1 & \hdots & F_0  & \hdots & 1 \\
m_1^2 & m_2^2 & \hdots & 0 &\hdots & m_m^2 \\
m_1^4 & m_2^4 & \hdots & 0 & \hdots & m_m^4 \\
\hdotsfor 6 \\
m_1^{2(m-2)} & m_2^{2(m-2)} & \hdots & 0 & \hdots &  m_m^{2(m-2)} \\
m_1^{2(m-1)} & m_2^{2(m-1)} & \hdots & 0 & \hdots & m_m^{2(m-1)} \\
\end{vmatrix}\qquad \label{ro11}\\
& & - \sum_{k=m+1}^n a_k
\begin{vmatrix}
1 & 1 & \hdots & 1 & \hdots & 1 \\
m_1^2 & m_2^2 & \hdots & m_k^2 &\hdots & m_m^2 \\
m_1^4 & m_2^4 & \hdots & m_k^4 & \hdots & m_m^4 \\
\hdotsfor 6 \\
m_1^{2(m-2)} & m_2^{2(m-2)} & \hdots & m_k^{2(m-2)} & \hdots &  m_m^{2(m-2)} \\
m_1^{2(m-1)} & m_2^{2(m-1)} & \hdots & m_k^{2(m-1)} & \hdots & m_m^{2(m-1)} \\
\end{vmatrix}\qquad \nonumber
\end{eqnarray}
from where, if in the first determinant the Laplace expansion by the entries
of the column $i$ is used, the explicit form is obtained
\begin{eqnarray}
det\mathbf{M_i}&=& F_0 (-1)^{1+i}\prod^m_{\substack{j=1\\j\neq
i}}m_j^2\prod^m_{\substack{j,l=1\\j<l,j,l\neq i}}(m_l^2-m_j^2)- \label{ro12} \\
&-&(-1)^{i-1}\prod^m_{\substack{j,l=1\\j<l, j,l\neq i}}(m_l^2-m_j^2)\sum_{k=m+1}^na_k\prod^m_{\substack{j=1\\j\neq
i}}(m_j^2-m_k^2).\nonumber
\end{eqnarray}
Now, substituting (\ref{ro8}) and (\ref{ro12}) into (\ref{ro9}),
one gets the solutions of (\ref{ro5}) as follows
\begin{eqnarray}
a_i&=&\frac{F_0(-1)^{1+i}\prod^m_{\substack{j=1\\j\neq i}}m_j^2\prod^m_{\substack{j,l=1\\j<l,j,l\neq
i}}(m_l^2-m_j^2)}{\prod^m_{\substack{j,l=1\\j<l}}(m_l^2-m_j^2)}-  \label{ro13} \\
&-&\frac{(-1)^{i-1}\prod^m_{\substack{j,l=1\\j<l,j,l\neq
i}}(m_l^2-m_j^2)\sum_{k=m+1}^na_k\prod^m_{\substack{j=1\\j\neq
i}}(m_j^2-m_k^2)}{\prod^m_{\substack{j,l=1\\j<l}}(m_l^2-m_j^2)}.
\nonumber
\end{eqnarray}

In order to find, by means of (\ref{ro13}), an explicit form of
$F_h(t)$ to be automatically normalized  with the required
asymptotic behaviour, let us separate the sum in (\ref{ro1}) into
two parts with the subsequent transformation of the first one into
a common denominator as follows:
\begin{eqnarray}
F_h(t)&=& \sum_{i=1}^m\frac{m_i^2a_i}{m_i^2-t} +
\sum_{k=m+1}^n\frac{m_k^2a_k}{m_k^2-t}= \label{ro14} \\
&=& \frac{\sum_{i=1}^m\prod^m_{\substack{j=1\\j\neq
i}}(m_j^2-t)m_i^2a_i}{\prod^m_{j=1}(m_j^2-t)} +
\sum^n_{k=m+1}\frac{m_k^2a_k}{m_k^2-t}. \nonumber
\end{eqnarray}

Then (\ref{ro13}) together with (\ref{ro14}) gives
\begin{eqnarray}
& &F_h(t)=F_0\frac{\sum_{i=1}^m(-1)^{1+i}m_i^2\prod^m_{\substack{j=1\\j\neq
i}}m_j^2\prod^m_{\substack{j=1\\j\neq
i}}(m_j^2-t)\prod^m_{\substack{j,l=1\\j<l, j,l\neq
i}}(m_l^2-m_j^2)}{\prod^m_{j=1}(m_j^2-t)\prod^m_{\substack{j,l=1\\j<l}}(m_l^2-m_j^2)}-\nonumber
\\
&-& \frac{\sum_{i=1}^m(-1)^{i-1}m_i^2\prod^m_{\substack{j=1\\j\neq
i}}(m_j^2-t)\prod^m_{\substack{j,l=1\\j<l, j,l\neq
i}}(m_l^2-m_j^2)\sum_{k=m+1}^na_k\prod^m_{\substack{j=1\\j\neq
i}}(m_j^2-m_k^2)}{\prod^m_{j=1}(m_j^2-t)\prod^m_{\substack{j,l=1\\j<l}}(m_l^2-m_j^2)}+
\nonumber \\
&+& \sum_{k=m+1}^n\frac{m_k^2a_k}{m_k^2-t}. \label{ro15}
\end{eqnarray}
The first term in (\ref{ro15}) can be rearranged into the form
\begin{equation}
F_0\frac{\prod^m_{j=1}m_j^2}{\prod^m_{j=1}(m_j^2-t)}\frac{\sum_{i=1}^m(-1)^{1+i}
\prod^m_{\substack{j=1\\j\neq
i}}(m_j^2-t)\prod^m_{\substack{j,l=1\\j<l, j,l\neq
i}}(m_l^2-m_j^2)}{\prod^m_{\substack{j,l=1\\j<l}}(m_l^2-m_j^2)}\label{ro16}
\end{equation}
in which one can prove explicitly the identity
\begin{equation}
\sum_{i=1}^m(-1)^{1+i}\prod^m_{\substack{j=1\\j\neq
i}}(m_j^2-t)\prod^m_{\substack{j,l=1\\j<l, j,l\neq
i}}(m_l^2-m_j^2)\equiv \prod^m_{\substack{j,l=1\\j<l}}(m_l^2-m_j^2)
\label{ro17}
\end{equation}
leading to remarkable simplification of the term under
consideration as follows:
\begin{equation}
F_0\frac{\prod^m_{j=1}m_j^2}{\prod^m_{j=1}(m_j^2-t)}.
\label{ro18}
\end{equation}

One could prove (\ref{ro17}) by rewriting its left-hand side into the
following form
\begin{eqnarray}
& &\sum_{i=1}^m(-1)^{1+i}\times \label{ro19} \\
& &\begin{vmatrix}
(m_1^2-t)   \hdots &(m_{i-1}^2-t)  & (m_{i+1}^2-t) \hdots & (m_m^2-t) \\
m_1^2(m_1^2-t)   \hdots & m_{i-1}^2(m_{i-1}^2-t) & m_{i+1}^2(m_{i+1}^2-t) \hdots & m_m^2(m_m^2-t) \\
m_{1}^4(m_{1}^2-t)   \hdots & m_{i-1}^4(m_{i-1}^2-t) & m_{i+1}^4(m_{i+1}^2-t) \hdots & m_m^4(m_{m}^2-t)\\
\hdotsfor 4 \\
m_1^{2(m-3)}(m_{1}^2-t)   \hdots & m_{i-1}^{2(m-3)}(m_{i-1}^2-t) &m_{i+1}^{2(m-3)}(m_{i+1}^2-t) \hdots &  m_m^{2(m-3)}(m_{m}^2-t) \\
m_1^{2(m-2)}(m_{1}^2-t)   \hdots & m_{i-1}^{2(m-2)}(m_{i-1}^2-t) & m_{i+1}^{2(m-2)}(m_{i+1}^2-t) \hdots & m_m^{2(m-2)}(m_{m}^2-t) \\
\end{vmatrix}\qquad \nonumber
\end{eqnarray}
and then by using various basic properties of the determinants
decomposing it into the sum of large number of various
determinants of the same order with their subsequent explicit
calculations. Since this procedure seems to be, from the
calculational point of view, not simple, with the aim of a proving
(\ref{ro17}) let us define  the new matrix
\begin{equation}
\mathbf{D(t)}= \begin{pmatrix}
1& 1  \hdots & 1& \hdots & 1\\
(m_1^2-t) & (m_2^2-t)   \hdots &(m_{i}^2-t)  & \hdots & (m_m^2-t) \\
(m_1^2-t)^2  & (m_2^2-t)^2 \hdots & (m_{i}^2-t)^2 & \hdots & (m_m^2-t)^2 \\
(m_{1}^2-t)^3  & (m_2^2-t)^3\hdots & (m_{i}^2-t)^3 & \hdots & (m_{m}^2-t)^3\\
\hdotsfor 5 \\
(m_{1}^2-t^{m-2} & (m_2^2-t)^{m-2} \hdots & (m_{i}^2-t)^{m-2} & \hdots & (m_{m}^2-t)^{m-2} \\
(m_{1}^2-t)^{m-1} & (m_2^2-t)^{m-1} \hdots & (m_{i}^2-t)^{m-1} &\hdots & (m_{m}^2-t)^{m-1} \\
\end{pmatrix}.\qquad\label{ro20}
\end{equation}
Denoting $(m_i^2-t)$=$x_i$ one gets the Vandermonde matrix
\begin{equation}
\mathbf{D(t)}= \begin{pmatrix}
1& 1  \hdots & 1 & \hdots & 1\\
x_1 & x_2   \hdots &x_{i}  & \hdots & x_m \\
x_1^2  & x_2^2 \hdots & x_{i}^2 & \hdots & x_m^2 \\
x_{1}^3  & x_2^3\hdots & x_{i}^3 & \hdots & x_m^3\\
\hdotsfor 5 \\
x_{1}^{m-2} & x_2^{m-2} \hdots & x_{i}^{m-2} & \hdots & x_{m}^{m-2} \\
x_{1}^{m-1} & x_2^{m-1} \hdots & x_{i}^{m-1} &\hdots & x_{m}^{m-1} \\
\end{pmatrix}\qquad \label{ro21}
\end{equation}
the determinant of which is equal just to the right-hand side of
(\ref{ro17})
\begin{equation}
det\mathbf{D(t)}=\prod^m_{\substack{j,l=1\\j<l}}(x_l-x_j)\equiv
\prod^m_{\substack{j,l=1\\j<l}}(m_l^2-t-m_j^2+t)=\prod^m_{\substack{j,l=1\\j<l}}(m_l^2-m_j^2).
\label{ro22}
\end{equation}

On the other hand, if in the determinant of the matrix
(\ref{ro20}) the Laplace expansion by the entries of the first row
with a subsequent pulling out of common factors in all columns of
the subdeterminants is carried out,  one gets the expression
\begin{eqnarray}
&&det\mathbf{D(t)}=\sum_{i=1}^m(-1)^{1+i}\prod^m_{\substack{j=1\\j\neq
i}}(m_j^2-t)\times  \label{ro23}\\
& &\begin{vmatrix}
 1  &\hdots & 1 & 1  \hdots & 1\\
(m_1^2-t) &   \hdots &(m_{i-1}^2-t)  & (m_{i+1}^2-t)\hdots & (m_m^2-t) \\
(m_1^2-t)^2  &  \hdots & (m_{i-1}^2-t)^2 &(m_{i+1}^2-t)^2 \hdots & (m_m^2-t)^2 \\
(m_{1}^2-t)^3  & \hdots & (m_{i-1}^2-t)^3 &(m_{i+1}^2-t)^3 \hdots & (m_{m}^2-t)^3\\
\hdotsfor 5 \\
(m_{1}^2-t)^{m-3} &  \hdots & (m_{i-1}^2-t)^{m-3} & (m_{i+1}^2-t)^{m-3}\hdots & (m_{m}^2-t)^{m-3} \\
(m_{1}^2-t)^{m-2} &  \hdots & (m_{i-1}^2-t)^{m-2} &(m_{i+1}^2-t)^{m-2}\hdots & (m_{m}^2-t)^{m-2} \\
\end{vmatrix}.\qquad \nonumber
\end{eqnarray}

Then calculating explicitly the determinant in (\ref{ro23})  by
using again the denotation $x_k=(m_k^2-t)$ for
$k=1,...,i-1,i+1,..m$,  one finally obtains
\begin{equation}
det\mathbf{D(t)}=\sum_{i=1}^{m}(-1)^{1+i}\prod^m_{\substack{j=1\\j\neq
i}}(m_j^2-t)\prod^m_{\substack{j,l=1\\j<l,j,l\neq i}}(m_l^2-m_j^2)
\label{ro24}
\end{equation}
just the left-hand side of (\ref{ro17}) and in this way the identity under consideration
is clearly proved.

The second and third term in (\ref{ro15}), transforming them to a
common denominator, can be unified into one following term:
\begin{eqnarray}
& & \sum_{k=m+1}^n\biggl \{
\frac{m_k^2\prod^m_{j=1}(m_j^2-t)\prod^m_{\substack{j,l=1\\j<l}}(m_l^2-m_j^2)}
{(m_k^2-t)\prod_{j=1}^m(m_j^2-t)\prod_{\substack{j,l=1\\j<l}}^m(m_l^2-m_j^2)}+\label{ro25}
\\
&+&\frac{(m_k^2-t)\sum^m_{i=1}(-1)^im_i^2\prod^m_{\substack{j=1\\j\neq
i}}(m_j^2-m_k^2)\prod^m_{\substack{j,l=1\\j<l,j,l\neq
i}}(m_l^2-m_j^2)\prod^m_{\substack{j=1\\j\neq
i}}(m_j^2-t)}{(m_k^2-t)\prod_{j=1}^m(m_j^2-t)\prod^m_{\substack{j,l=1\\j<l}}(m_l^2-m_j^2)}\biggr
 \}a_k \nonumber
\end{eqnarray}
the numerator of which is exactly the Laplace  expansion by the
entries of the first row of the determinant of the matrix of the  $(m+1)$
order
\begin{equation}
\mathbf{N(t)}= \begin{pmatrix}
m_k^2 & m_1^2  & \hdots  & m_m^2\\
(m_k^2-t) & (m_1^2-t)      & \hdots & (m_m^2-t) \\
(m_k^2-t)^2  & (m_1^1-t)^2   & \hdots & (m_m^2-t)^2 \\
(m_k^2-t)^3  & (m_1^2-t)^3  & \hdots & (m_{m}^2-t)^3\\
\hdotsfor 4 \\
(m_k^2-t)^{m-1} & (m_1^2-t)^{m-1}   & \hdots & (m_{m}^2-t)^{m-1} \\
(m_k^2-t)^{m} & (m_1^2-t)^{m}  &\hdots & (m_{m}^2-t)^{m} \\
\end{pmatrix}.\qquad \label{ro26}
\end{equation}
If we define the new matrix of the $(m+1)$ order
\begin{equation}
\mathbf{R(t)}= \begin{pmatrix}
1& 1 & \hdots  & 1\\
(m_k^2-t) & (m_1^2-t)    & \hdots & (m_m^2-t) \\
(m_k^2-t)^2  & (m_1^2-t)^2  & \hdots & (m_m^2-t)^2 \\
(m_k^2-t)^3  & (m_1^2-t)^3  & \hdots & (m_{m}^2-t)^3\\
\hdotsfor 4 \\
(m_k^2-t)^{m-1} & (m_1^2-t)^{m-1}   & \hdots & (m_{m}^2-t)^{m-1} \\
(m_k^2-t)^{m} & (m_1^2-t)^{m}  &\hdots & (m_{m}^2-t)^{m} \\
\end{pmatrix},\qquad \label{ro27}
\end{equation}
then for the determinant of both matrices, (\ref{ro26}) and
(\ref{ro27}), the equation
\begin{equation}
det\mathbf{N(t)}- t\cdot det\mathbf{R(t)}\equiv det \mathbf{S(t)}=0
\label{ro28}
\end{equation}
is fulfilled under the assumption  that $det\mathbf{S(t)}$ is
obtained by   multiplication of the first row of
$det\mathbf{R(t)}$ by $t$, and the substraction of the resultant
determinant from $\mathbf{N(t)}$ is carried out explicitly.

There is valid also a relation
\begin{equation}
det\mathbf{N(0)}=0 \label{ro29}
\end{equation}
as in $det\mathbf{N(0)}$ (like in $det\mathbf{S(t)}$) the first two
rows are identical.

Now, in order to arrange the numerator of (\ref{ro25})
cenveniently, we write  $det\mathbf{N(t)}$ in the form
\begin{equation}
det\mathbf{N(t)}=t\cdot det\mathbf{R(0)}-det\mathbf{N(0)},
\label{ro30} \end{equation} taking into account (\ref{ro28}),
(\ref{ro29}) and the identity
\begin{equation} det\mathbf{R(t)}\equiv det \mathbf{R(0)}.
\label{ro31}
\end{equation}
In (\ref{ro30}) we apply the Laplace expansion by entries of the
first row to $det\mathbf{R(0)}$ and $det\mathbf{N(0)}$, separately.
As a result, one gets
\begin{eqnarray}
& & det\mathbf{N(t)}=t\cdot \begin{vmatrix}
m_1^2 & m_2^2 &  \hdots & m_m^2 \\
m_1^4 & m_2^4  & \hdots & m_m^4 \\
\hdotsfor 4 \\
m_1^{2(m-1)} & m_2^{2(m-1)} & \hdots &   m_m^{2(m-1)} \\
m_1^{2m} & m_2^{2m} & \hdots  & m_m^{2m} \\
\end{vmatrix}\qquad + \label{ro32}\\
&+&\sum_{i=1}^m(-1)^it\begin{vmatrix}
m_k^2 & m_1^2  \hdots & m_{i-1}^2& m_{i+1}^2  & \hdots & m_m^2 \\
m_k^4 & m_1^4  \hdots & m_{i-1}^4 & m_{i+1}^4 &\hdots & m_m^4 \\
\hdotsfor 6 \\
m_k^{2(m-1)} & m_1^{2(m-1)}   \hdots & m_{i-1}^{2(m-1)} & m_{i+1}^{2(m-1)} & \hdots &  m_m^{2(m-1)} \\
m_k^{2m} & m_1^{2m}  \hdots &m_{i-1}^{2m} & m_{i+1}^{2m}  & \hdots & m_m^{2m} \\
\end{vmatrix}\qquad -\nonumber \\
&-& m_k^2 \begin{vmatrix}
m_1^2 & m_2^2 &  \hdots & m_m^2 \\
m_1^4 & m_2^4   &\hdots & m_m^4 \\
\hdotsfor 4 \\
m_1^{2(m-1)} & m_2^{2(m-1)} & \hdots &   m_m^{2(m-1)} \\
m_1^{2m} & m_2^{2m} & \hdots  & m_m^{2m} \\
\end{vmatrix}\qquad -\nonumber \\
&-&\sum_{i=1}^m(-1)^im_i^2\begin{vmatrix}
m_k^2 & m_1^2  \hdots & m_{i-1}^2& m_{i+1}^2  & \hdots & m_m^2 \\
m_k^4 & m_1^4  \hdots & m_{i-1}^4 & m_{i+1}^4 &\hdots & m_m^4 \\
\hdotsfor 6 \\
m_k^{2(m-1)} & m_1^{2(m-1)}   \hdots & m_{i-1}^{2(m-1)} & m_{i+1}^{2(m-1)} & \hdots &  m_m^{2(m-1)} \\
m_k^{2m} & m_1^{2m}  \hdots &m_{i-1}^{2m} & m_{i+1}^{2m}  & \hdots & m_m^{2m} \\
\end{vmatrix}\qquad \nonumber
\end{eqnarray}

or calculating explicitly the corresponding subdeterminants
\begin{eqnarray}
& &det \mathbf{N(t)}=
(t-m_k^2)\prod_{j=1}^mm_j^2\prod^m_{\substack{j,l=1\\j<l}}(m_l^2-m_j^2)+
\label{ro33} \\
&+& \sum_{i=1}^m(-1)^i(t-m_i^2)m_k^2\prod^m_{\substack{j=1\\
j\neq i}}m_j^2\prod^m_{\substack{j=1\\j\neq
i}}(m_j^2-m_k^2)\prod^m_{\substack{j,l=1\\j<l,j,l\neq
i}}(m_l^2-m_j^2).\nonumber
\end{eqnarray}
Substituting the latter into (\ref{ro25}) one obtains
\begin{equation}
\sum_{k=m+1}^n\biggl \{ -\frac{\prod_{j=1}^m
m_j^2}{\prod_{j=1}^m(m_j^2-t)}
+\sum_{i=1}^m\frac{m_k^2}{(m_k^2-t)}\frac{\prod_{\substack{j=1\\j\neq
i}}^mm_j^2}{\prod^m_{\substack{j=1\\j\neq
i}}(m_j^2-t)}\frac{\prod^m_{\substack{j=1\\j\neq
i}}(m_j^2-m_k^2)}{\prod^m_{\substack{j=1\\j\neq
i}}(m_j^2-m_i^2)}\biggr \}a_k \label{ro34}
\end{equation}
and combining  this result with (\ref{ro18}), one gets the form
factor $F_h(t)$ to be saturated by $n$-vector mesons $(n>m)$ in
the form suitable for the unitarization
\begin{eqnarray}
& &F_h(t)=F_0\frac{\prod^m_{j=1}m_j^2}{\prod^m_{j=1}(m_j^2-t)}+ \label{ro35} \\
&+& \sum_{k=m+1}^n\biggl\{\sum_{i=1}^m\frac{m_k^2}{(m_k^2-t)}
\frac{\prod_{\substack{j=1\\j\neq
i}}^mm_j^2}{\prod_{\substack{j=1\\j\neq
i}}^m(m_j^2-t)}\frac{\prod^m_{\substack{j=1\\j\neq
i}}(m_j^2-m_k^2)}{\prod^m_{\substack{j=1\\j\neq
i}}(m_j^2-m_i^2)}-\frac{\prod_{j=1}^mm_j^2}{\prod_{j=1}^m(m_j^2-t)}\biggr
\}a_k \nonumber
\end{eqnarray}
for which the asymptotic behaviour (\ref{ro2}) and for $t=0$ the
normalization (\ref{ro3}) are fulfilled automatically.

The asymptotic behaviour in (\ref{ro35}) is transparent. However,
for the normalization (\ref{ro3}) the following identity:
\begin{equation}
\sum_{i=1}^m\frac{\prod^m_{\substack{j=1\\j\neq
i}}(m_j^2-m_k^2)}{\prod^m_{\substack{j=1\\j\neq
i}}(m_j^2-m_i^2)}=1 \label{ro36}
\end{equation}
has to be valid in the second term of (\ref{ro35}) generally.

For $m=2,3,4,5$ it can be proved explicitly. And for an arbitrary
finite $m$ it follows directly from (\ref{ro25}), the numerator of
which is exactly the Laplace expansion by the entries of the first
row of the  determinant of the matrix (\ref{ro26}). Then, just
relation (\ref{ro29}) causes the term (\ref{ro25}) and also
(\ref{ro34}) at $t=0$ for arbitrary nonzero values of $a_k$ to be
zero. Hence, every term in the wave-brackets of (\ref{ro34}) for $t=0$
has to be zero and this is true if and only if  identity
(\ref{ro36}) is fulfilled.

Now we consider the case of equations (\ref{ro4}) for $n=m$. Then
it can  also be rewritten into the matrix form (\ref{ro5}) with the  $m\times m$
Vandermonde matrix (\ref{ro6}) and the same
column vector $\mathbf{a}$, but with the $\mathbf{b}$ vector of
the following form: \begin{equation}
\mathbf{b}=\begin{pmatrix} F_0\\0 \\0 \\
\vdots\\ 0 \\0 \\ \end{pmatrix}.\qquad \label{ro37}
\end{equation}
So  the corresponding solutions are again looked for in the form
$$
a_i=\frac{det\mathbf{M_i}}{det\mathbf{M}},
$$
but with the matrix $\mathbf{M_i}$
\begin{equation}
\mathbf{M_i}=\begin{pmatrix} 1 &  \hdots & 1 & F_0  & 1 & \hdots &
1\\ m_1^2 &  \hdots & m_{i-1}^2 & 0 & m_{i+1}^2 & \hdots &
m_m^2
\\ m_1^4 & \hdots & m_{i-1}^4 & 0 & m_{i+1}^4  & \hdots & m_m^4 \\ \hdotsfor 7\\
m_1^{2(m-2)} &  \hdots  & m_{i-1}^{2(m-2)} &  0 & m_{i+1}^{2(m-2)} & \hdots & m_m^{2(m-2)} \\
m_1^{2(m-1)} &  \hdots & m_{i-1}^{2(m-1)} & 0 & m_{i+1}^{2(m-1)} & \hdots &  m_m^{2(m-1)} \\
\end{pmatrix},\qquad           \label{ro38}
\end{equation}
and as result
\begin{equation}
det\mathbf{M_i}=F_0(-1)^{1+i}\prod^m_{\substack{j=1\\j\neq
i}}m_j^2\prod^m_{\substack{j,l=1\\j<l, j,l\neq i}}(m_l^2-m_j^2)
\label{ro39}
\end{equation}
and the solutions
\begin{eqnarray}
a_i&=&F_0 \frac{(-1)^{1+i}\prod^m_{\substack{j=1\\j\neq
i}}m_j^2\prod^m_{\substack{j,l=1\\j<l, j,l\neq
i}}(m_l^2-m_j^2)}{\prod^m_{\substack{j,l=1\\j<l}}(m_l^2-m_j^2)}=
\label{ro40}\\
&=& F_0\frac{\prod^m_{\substack{j=1\\j\neq
i}}m_j^2(-1)^{1+i}}{\prod^m_{\substack{j=1\\j\neq
i}}(m_j^2-m_i^2)(-1)^{i-1}}\nonumber
\end{eqnarray}
are completely expressed only through the masses of $m$
vector-mesons  by means of which the considered FF is saturated.

The third case with the $(m-1)$ linear homogeneous algebraic equations
for the $n$ $\;\;$ $(n>m)$ coupling constant ratios without any
normalization of FF appears naturally in the  determination of
strange FF behaviours of a strongly interacting particles with the
spin $s>0$ from the isoscalar parts of the corresponding EM FF's,
as we have mentioned in Introduction.

Then, we have only the equations \begin{equation}
\sum_{i=1}^nm_i^{2r}a_i = 0, \;\;\; r=1,2,...m-1 \label{ro41}
\end{equation}
which can be rewritten in the matrix form (\ref{ro5}) with the
$(m-1)\times(m-1)$ matrix $\mathbf{M}$
\begin{equation}
\mathbf{M}=\begin{pmatrix}
m_1^2 & m_2^2 & \hdots & m_{m-1}^2 \\
m_1^4 & m_2^4 & \hdots & m_{m-1}^4 \\
\hdotsfor 4 \\
m_1^{2(m-1)} & m_2^{2(m-1)} & \hdots &  m_{m-1}^{2(m-1)} \\
\end{pmatrix}\qquad                                        \label{ro42}
\end{equation}
and the column vectors
\begin{eqnarray}
\mathbf{a}=\begin{pmatrix} a_1\\a_2 \\a_3 \\
\vdots\\a_{m-1}\\\end{pmatrix},\qquad \;\;\;
\mathbf{b}=\begin{pmatrix}
-\sum_{k=m}^nm_k^2a_k\\-\sum_{k=m}^nm^4_k a_k \\
-\sum_{k=m}^nm^6_k a_k
\\ \hdotsfor 1\\ -\sum_{k=m}^nm^{2(m-1)}_k a_k\\
\end{pmatrix}. \qquad
\label{ro43}
\end{eqnarray}
The determinant of the matrix $\mathbf{M}$
\begin{equation}
det\mathbf{M}=\prod^{m-1}_{j=1}
m_j^2\prod^{m-1}_{\substack{j,l=1\\j<l}}(m_l^2-m_j^2) \label{ro44}
\end{equation}
is different from zero, and thus, a nontrivial solution of
(\ref{ro41}) exists
\begin{equation*}
a_i=\frac{det\mathbf{M_i}}{det\mathbf{M}}
\end{equation*}
where the matrix $\mathbf{M_i}$ takes the form
\begin{equation}
\mathbf{M_i}=\begin{pmatrix}
 m_1^2 & \hdots & m_{i-1}^2 &
-\sum_{k=m}^nm_k^2a_k & m_{i+1}^2 & \hdots & m_{m-1}^2
\\ m_1^4 & \hdots & m_{i-1}^4 & -\sum_{k=m}^nm_k^4a_k& m_{i+1}^4  & \hdots & m_{m-1}^4 \\ \hdotsfor 7\\
 m_1^{2(m-1)} &  \hdots & m_{i-1}^{2(m-1)}  & -\sum_{k=m}^nm_k^{2(m-1)}a_k & m_{i+1}^{2(m-1)} & \hdots &  m_m^{2(m-1)} \\
\end{pmatrix}.\qquad           \label{ro45}
\end{equation}
Then employing the basic properties of the determinants one gets
\begin{eqnarray}
&&det\mathbf{M_i}=-\sum_{k=m}^na_k
\begin{vmatrix}
 m_1^2 & \hdots & m_{i-1}^2 &
m_k^2 & m_{i+1}^2 & \hdots & m_{m-1}^2\\
 m_1^4 & \hdots & m_{i-1}^4 & m_k^4 & m_{i+1}^4  & \hdots & m_{m-1}^4 \\
 m_1^6 & \hdots & m_{i-1}^6 & m_k^6 & m_{i+1}^6  & \hdots & m_{m-1}^6 \\
\hdotsfor 7\\
 m_1^{2(m-1)} &  \hdots & m_{i-1}^{2(m-1)}  & m_k^{2(m-1)} & m_{i+1}^{2(m-1)} & \hdots &  m_{m-1}^{2(m-1)} \\
\end{vmatrix}=\qquad           \nonumber \\
&=&-\sum_{k=m}^na_km_k^2\prod^{m-1}_{\substack{j=1\\j\neq i}}m_j^2
\begin{vmatrix} 1& \hdots& 1 & 1& 1&\hdots & 1\\
 m_1^2 & \hdots & m_{i-1}^2 &
m_k^2 & m_{i+1}^2 & \hdots & m_{m-1}^2
\\ m_1^4 & \hdots & m_{i-1}^4 & m_k^4 & m_{i+1}^4  & \hdots & m_{m-1}^4 \\
\hdotsfor 7\\
 m_1^{2(m-2)} &  \hdots & m_{i-1}^{2(m-2)}  & m_k^{2(m-2)} & m_{i+1}^{2(m-2)} & \hdots &  m_m^{2(m-2)} \\
\end{vmatrix}=\qquad          \nonumber \\
&=&-\sum_{k=m}^n m_k^2a_k\prod^{m-1}_{\substack{j=1\\j\neq i}}m_j^2
\prod^{m-1}_{\substack{j,l=1\\j<l, j,l\neq i}}(m_l^2-m_j^2)(-1)^{i-1}
\prod^{m-1}_{\substack{j=1\\j\neq i}}(m_j^2-m_k^2).\label{ro46}
\end{eqnarray}
As a result,
\begin{equation}
a_i=\frac{-\sum^n_{k=m} m_k^2a_k\prod^{m-1}_{\substack{j=1\\j \neq
i}}m_j^2\prod^{m-1}_{\substack{j,l=1\\j<l, j,l\neq
i}}(m_l^2-m_j^2)(-1)^{i-1}\prod^{m-1}_{\substack{j=1\\j\neq
i}}(m_j^2-m_k^2)}{\prod^{m-1}_{j=1}m_j^2\prod^{m-1}_{\substack{j,l=1\\j<l}}(m_j^2-m_l^2)}
\label{ro47}
\end{equation}
or finally,
\begin{equation}
a_i=-\sum_{k=m}^n\frac{m_k^2}{m_i^2}\frac{\prod^{m-1}_{\substack{j=1\\j
\neq i}}(m_j^2-m_k^2)}{\prod^{m-1}_{\substack{j=1\\ j\neq
i}}(m_j^2-m_i^2)}a_k, \;\;\; i=1,2...,m-1. \label{ro48}
\end{equation}
Now  substituting (\ref{ro47}) into
\begin{equation}
F_h(t)=\frac{\sum^{m-1}_{i=1}\prod^{m-1}_{\substack{j=1\\j\neq
i}}(m_j^2-t)m^2_ia_i}{\prod_{j=1}^{m-1}(m_j^2-t)} +
\sum_{k=m}^n\frac{m_k^2a_k}{m_k^2-t} \label{ro49}
\end{equation}
and transforming both terms into a common denominator one gets the
relation
\begin{eqnarray}
&&F_h(t)=\sum_{k=m}^n\biggl\{\frac{\prod^{m-1}_{j=1}(m_j^2-t)\prod^{m-1}_{\substack{j,l=1\\j<l}}(m^2_l-m^2_j)}
{\prod^{m-1}_{j=1}(m_j^2-t)\prod^{m-1}_{\substack{j,l=1\\j<l}}(m^2_l-m^2_j)}+
\label{ro50} \\
&+&\frac{(m_k^2-t)\sum_{i=1}^{m-1}(-1)^i\prod^{m-1}_{\substack{j=1\\j\neq
i}}(m_j^2-t)\prod^{m-1}_{\substack{j,l=1\\j<l, j,l\neq
i}}(m_l^2-m_j^2)\prod^{m-1}_{\substack{j=1\\j\neq
i}}(m_j^2-m_k^2)}{\prod^{m-1}_{j=1}(m_j^2-t)\prod^{m-1}_{\substack{j,l=1\\j<l}}(m^2_l-m^2_j)}\biggr\}\times\nonumber
\\
 & &\times\frac{m_k^2}{m_k^2-t}a_k, \nonumber
\end{eqnarray}
in which the numerator of the first term under the sum is just the
Laplace expansion by the entries of the first row of the determinant
of the matrix $\mathbf{T(t)}$ of the  $m$ order
\begin{equation}
\mathbf{T(t)}= \begin{pmatrix}
1& 1   &\hdots  & 1\\
(m_k^2-t) & (m_1^2-t)   &  \hdots   &  (m_{m-1}^2-t) \\
(m_k^2-t)^2  & (m_1^2-t)^2   & \hdots & (m_{m-1}^2-t)^2 \\
\hdotsfor 4 \\
(m_{k}^2-t)^{m-1} & (m_1^2-t)^{m-1} & \hdots  & (m_{m-1}^2-t)^{m-1} \\
\end{pmatrix}.\qquad\label{ro51}
\end{equation}
One can  immediately prove that
\begin{equation}
det \mathbf{T(t)}\equiv det \mathbf{T(0)}. \label{ro52}
\end{equation}
Then calculating $det\mathbf{T(0)}$ explicitly
\begin{equation}
det\mathbf{T(0)}=\prod^{m-1}_{j=1}(m_j^2-m_k^2)\prod^{m-1}_{\substack{j,l=1\\j<l}}(m_l^2-m_j^2)
\label{ro53}
\end{equation}
and substituting the result into (\ref{ro50}) instead of the
numerator of the term in the wave-brackets, one finally obtains the
parametrization
\begin{equation}
F_h(t)=\sum_{k=m}^n\frac{\prod^{m-1}_{j=1}(m_j^2-m_k^2)}{\prod_{j=1}^{m-1}m_j^2}\frac{\prod^{m-1}_{j=1}m_j^2}
{\prod_{j=1}^{m-1}(m_j^2-t)}\frac{m_k^2}{m_k^2-t}a_k  \label{ro54}
\end{equation}
for which the asymptotic behaviour (\ref{ro2}) is fulfilled
automatically.

\section{Conclusions}

General solutions of asymptotic conditions for EM FF's of hadrons
represented by the VMD model, derived by means of the superconvergent
sum rules for the imaginary part of FF under consideration, in
which the coefficients are simply even powers of the corresponding
vector-meson masses, have been found.

We have distinguished three cases appearing in various physical
situations:
\begin{itemize}
\item[i)] in the construction of  unitary and analytic models
of the EM structure of any strongly interacting particle with a
number of building quarks $n_q>2$  when at the first stage one has
need for the VMD parametrization of FF to be saturated with $n$
different vector mesons, but with the required asymptotics (\ref{ro2})
and normalization (\ref{ro3}), under the assumption $m<n$
\item[ii)] in the same problem, however, when $m=n$
\item[iii)]
in the  determination of the strange form factor behaviours of a strongly
interacting particle from the isoscalar parts of the corresponding
electromagnetic form factors. \end{itemize}

In the first case, we have found the explicit form (\ref{ro35}) of
EM FF for which the asymptotic behaviour (\ref{ro2}) and for $t=0$
the normalization (\ref{ro3}) are fulfilled automatically. Such a
form is the starting point in the  construction of the unitary and
analytic model of the EM structure of any strongly interacting
particle  in which a superposition of vector-meson poles and
continuum contributions are considered at the same time.

In the second case, the explicit expressions (\ref{ro40}) of all
considered coupling constant ratios are found to be expressed
through the masses of saturated vector mesons and the norm $F_0$
of FF. The direct application of (\ref{ro40})  to nucleons
\cite{dub5} gives surprising coincidence with the values obtained
by a fit \cite{dub6} of existing experimental data by means of the
modified VMD model.

In the third case, the explicit form (\ref{ro54}) of the isoscalar
part of  EM FF of the strongly interacting particle with the spin
$s>0$ was obtained,  which can be used to predict the behaviour of the strange
magnetic FF.

This work was  supported in part by the Slovak Grant Agency for
Sciences, Grant No. 2/1111/21 (S.~D.) and Grant No. 1/7068/21
(A.~Z.~D).


\begin{thebibliography}{99}

\bibitem{dub1}
C.~Adamu\v s\v cin, A.~Z.~Dubni\v ckov\'a, S.~Dubni\v cka,
R.~Pek\'arik, P.~Weisenpacher, hep-ph/0203175.

\bibitem{dub2}
S.~Dubni\v cka, Acta Physica Polonica {\bf B27},  2525 (1996).

\bibitem{dub3}
S.~Dubni\v cka, A.~Z.~Dubni\v ckov\'a, E.~Krn\'a\v c, Phys. Lett.
{\bf B261}, 127 (1991).

\bibitem{dub4}
S.~Dubni\v cka, A.~Z.~Dubni\v ckov\'a, P.~Weisenpacher,
hep-ph/0102171.

\bibitem{dub5}
D.~Drechsel, J.~Becker, A.~Z.~Dubni\v ckov\'a, S.~Dubni\v cka et
al.: Hadron polarizabilities and form factors. Lecture Notes in Physics No. 513, Chiral Dynamics: Theory and experiment. Eds.:
A.M. Bernstein, D. Drechsel, Th. Walcher, Springer Verlag, Heidelberg (1998) p. 264-291.

\bibitem{dub6}
P.~Mergel, Ulf-G. Meissner, D. Drechsel, Nucl. Phys. {\bf A596},
367 (1996).

\end{thebibliography}
\end{document}